\documentstyle[aaspp4,12pt,epsf]{article}

\begin{document}

\title{The Mid-Infrared Color-Luminosity Relation and the Local 12 $\mu$m Luminosity Function}

\author{Fan Fang, David L. Shupe, Cong Xu, Perry B. Hacking}
\affil{Infrared Processing and Analysis Center, Jet Propulsion Laboratory, 
Caltech 100-22, Pasadena, CA 91125}

\begin{abstract}

We have established a model to systematically estimate the contribution of
the mid-infrared emission features between 3 $\mu$m and 11.6 $\mu$m to the
IRAS in-band fluxes, using the results of ISO PHT-S observation of 16 galaxies
by Lu et al. (1997).  The model is used to estimate more properly the
$k$-corrections for calculating the restframe 12 and 25 $\mu$m fluxes
and luminosities of IRAS galaxies.

We have studied the 12-25 $\mu$m color-luminosity relation for a sample
of galaxies selected at 25 $\mu$m.
The color is found to correlate well with the 25 $\mu$m luminosity, the
mid-infrared luminosity, and the ratio of far-infrared and the blue
luminosities.  The relations with the mid-infrared luminosities are more
sensitive to different populations of galaxies, while a single relation of
the 12-25 $\mu$m color vs. the ratio of the far-infrared and
the blue luminosities applies equally well to these different populations.
The luminous and ultraluminous infrared galaxies
have redder 12-25 $\mu$m colors than those of the quasars.  These relations
provide powerful tools to differentiate different populations of galaxies.

The local luminosity function at 12 $\mu$m provides the basis
for interpreting the results of deep mid-infrared surveys planned or
in progress with ISO, WIRE and SIRTF.  We have selected a sample
of 668 galaxies from the IRAS Faint Source Survey flux-density limited
at 200 mJy at 12 $\mu$m.  A 12 $\mu$m local luminosity function is derived
and, for the first time in the literature, effects of density variation
in the local universe are considered and corrected in the calculation of the
12 $\mu$m luminosity function.  It is also found that the
12 $\mu$m-selected sample are dominated by quasars and active galaxies,
which therefore strongly affect the 12 $\mu$m luminosity function
at high luminosities.  The ultraluminous infrared galaxies are relatively
rare at 12 $\mu$m comparing with a 25 $\mu$m sample.

\end{abstract}


\section{Introduction}
\label{sec:intro}

The mid-infrared (MIR) spectral region is well-suited for studying
starburst and ultraluminous galaxies.  About 40\% of the luminosity
from starburst galaxies is radiated from 8-40 $\mu$m
(\markcite{soi87}Soifer et al.  1987).  Extinction effects are small,
and infrared cirrus emission is reduced at these wavelengths relative to
far-infrared bands.
For a fixed telescope aperture, the spatial resolution is also higher
at shorter wavelengths, and the confusion limit lies at higher redshifts.
All the recent and near-future infrared space missions, such as the
{\it Infrared Space Observatory (ISO)}, the {\it Wide-Field Infrared Explorer
(WIRE)}, and the {\it Space Infrared Telescope Facility (SIRTF)}
will conduct surveys in mid-infrared bands.  {\it WIRE}, a Small Explorer
mission due to launch in late 1998 (\markcite{hac96}Hacking et al.\ 1996;
\markcite{schemb96}Schember et al. 1996), will conduct a very deep survey
at 12 and 24 $\mu$m to study the evolution of starburst galaxies.
To interpret the results of these surveys now in
progress or soon to commence, it is necessary to better understand
the mid-infrared properties of galaxies in the local Universe.

One of the most important tools for extracting the rate and type of galaxy
evolution from a mid-infrared survey is the faint source counts.
A local mid-infrared luminosity
function is the basis for calculating the mid-infrared faint source counts
incorporating different evolutionary scenarios, and to extract the evolution
by comparing with observations.  Mid-infrared luminosity functions
have been calculated at 12 and 25 $\mu$m (Soifer \& Neugebauer 1991)
using a 60 $\mu$m selected IRAS sample (Soifer et al. 1987).  More recently,
\markcite{rush93}Rush et al. (1993) selected a 12 $\mu$m flux-limited sample
and calculated the luminosity functions for Seyfert and non-Seyfert
galaxies in the sample.  In a previous paper (\markcite{paper1}Shupe
et al. (1997), Paper I hereafter), we have presented the results of
a 25 $\mu$m luminosity function calculated from a large flux-limited
IRAS sample containing 1456 galaxies.  We continue to select
a flux-limited sample and calculate the luminosity function
at 12 $\mu$m in this paper.

The relation between the mid-infrared color and luminosity plays another
important role in estimating
various properties of galaxy evolution.  It defines distinct regions in the
color-flux diagram, for example, for different types of evolution and for
different populations of galaxies.  Such a relation is indicated by
the 12-25 $\mu$m vs. 60-100 $\mu$m color-color relation or by the 12-25 $\mu$m
color vs. the far-infrared luminosity relation obtained from
the IRAS survey (\markcite{soi91}Soifer \& Neugebauer 1991), and can
be estimated from a large sample of galaxies selected at mid-infrared bands.

Emission features near 12 $\mu$m thought to be produced by
aromatic hydrocarbon molecules have been observed in many astronomical spectra
(e.g., \markcite{gill73}Gillett et al. 1973; \markcite{russ78}Russell et al.
1978; \markcite{sell84}Sellgren 1984; \markcite{roch91}Roche, Aitken, \& Smith
1991; \markcite{oboul96}Boulade et al. 1996;
\markcite{vig96} Vigroux et al. 1996;
\markcite{met96}Metcalfe et al. 1996; \markcite{ces96}Cesarsky et al.
1996; \markcite{lu97}Lu et al. 1997).  These broad emission features
complicate the calculations of $k$-corrections and the fluxes and luminosities
at mid-infrared bands.  Fortunately, ISO observations have resulted in
high-quality mid-infrared spectra in various astronomical
circumstances, and the on-going surveys of IRAS galaxies using ISO
can provide an especially useful handle on this problem.

In the next section we present a model to systematically calculate
the contribution of the emission features in the mid-infrared bands
of IRAS galaxies.  The model is then incorporated in the following sections.
Section \ref{sec:clrlum} discusses the mid-infrared color-luminosity
relation obtained from the large 25 $\mu$m-selected sample of Paper I.
The population-dependency of the relation is discussed.  Then we
present the calculation of the 12 $\mu$m luminosity function
in Section \ref{sec:lumfcn}.  We first discuss a selection of galaxy sample
flux-limited at 12 $\mu$m from the IRAS Faint Source Survey in
Section \ref{sec:sample}.  Then the luminosity function is derived
and corrected for density variations in Section \ref{sec:pahlf}.
In Section \ref{sec:nopahlf} we discuss the effects of active galaxies
and quasars on the 12 $\mu$m luminosity function.
We summarize our results in Section \ref{sec:conclusion}.

\section{Calculating the flux of the mid-infrared emission features and the
$k$-correction}
\label{sec:pah}

The mid-infrared spectral energy distributions of galaxies contain
broad emission features, centered at
3.3 $\mu$m, 6.2 $\mu$m, 7.7 $\mu$m, 8.6 $\mu$m, and 11.3 $\mu$m, from very
small grains.  They have been interpreted as emission from aromatic
hydrocarbons molecules (\markcite{lp84}L\'{e}ger \& Puget 1984).
Since the first detection of the narrow emission band at 11.3 $\mu$m
by \markcite{gill73}Gillet et al. (1973),
the emission features have been found in many astronomical spectra, including
those of infrared-quiescent (e.g. Boulade et al. 1996) and starburst
(e.g. Vigroux et al. 1996) galaxies.  For more distant sources,
the emission features are redshifted into and out of the IRAS bands,
especially the 12 $\mu$m band.  They significantly change
the observed fluxes at these bands, and affect the $k$-correction
and the calculations of the rest-frame fluxes and luminosities.

At this stage, we want to establish a simple model which can estimate
the contribution of the emission to the IRAS in-band fluxes in a general
way.  Using the ISO PHT-S spectrometer, \markcite{lu97}Lu et al. (1997)
obtained the mid-infrared spectra of 16 nearby galaxies covering a range
of far-infrared luminosities.  Based on the data and a spectra template
kindly provided by George Helou and Nanyao Lu at IPAC, we model the
relation between the emission and the far-infrared color as a step function:
\begin{equation}
\label{eqn:pahrat}
\frac{\rm EF}{\rm FIR} = \left \{ \begin{array}{ll}
 0.12 & \mbox{if $F_{\nu}(60\mu m) < 0.6F_{\nu}(100\mu m)$} \\
 0.06 & \mbox{otherwise}
\end{array}
\right .,
\end{equation}
where EF and FIR refer to the flux due to the four emission features at
6.2 $\mu$m, 7.7 $\mu$m, 8.6 $\mu$m, and 11.3 $\mu$m (the 3.3 $\mu$m feature
is not clearly detected for most of the 16 galaxies), and
the far-infrared flux (defined as $1.26\times 10^{-14}(2.58f_{60\mu\rm m}
+f_{100\mu\rm m})\hspace{2mm} \rm watt \hspace{1mm} m^{-2}$), respectively,
and $F_{\nu}$ is the flux density.
The IRAS in-band relative flux of the emission features for
a given galaxy at a given redshift is calculated using the spectra template
and is then scaled according to the above relation to obtain the physical
emission flux.  The template spectrum we use will be published by Xu et al.
(1998).

Our model is based on such a relation with far-infrared flux and color
because, as shown by the results of Lu et al. (1997), the relative strength of
the emission features (with respect to the far-infrared flux) appears
to decrease with increasing starburst strength, characterized
by the 60-100 $\mu$m color in the model.  Physically, strong starburst
activities can destroy small grains, thereby reduce their emission.  
These grains would also not survive if, say, there is strong enough
activity in a galaxy nucleus.  Our empirical model
does not intend to discriminate between these phenomena.

The results by Lu et al. (1997) also show that the strength of the emission
features relative to the mid-infrared (3 to 11.6 $\mu$m) flux remains roughly
constant ($\sim 50\%$) from galaxy to galaxy.  This suggests that
the continuum component within this wavelength range correlates with
the emission features, and that they both have the same origin.
It is possible that a blend of weak features by similar grains form most of
the continuum (\markcite{mld96}Moutou, L\'{e}ger \& d'Hendecourt 1996).
The dominance of interstellar single-photon heating, which is typical
when radiation intensity does not exceed some critical value
(\markcite{fboul96}Boulanger et al. 1996), may have caused the general
similarity of the emission spectra of normal galaxies.
When radiation is strong enough, multiple-photon heating of small dust
particles can contribute more significantly to the 12 $\mu$m continuum,
as suggested by several ISO spectra of strong starburst regions (e.g. Vigroux
et al. 1996 for the Antennae galaxies; \markcite{lut97}Lutz et al. 1997 for M82).
The grains can be destroyed by the radiation, and the remaining emission features
can be swamped in the increased continuum in these regions.

\begin{figure}
\plotone{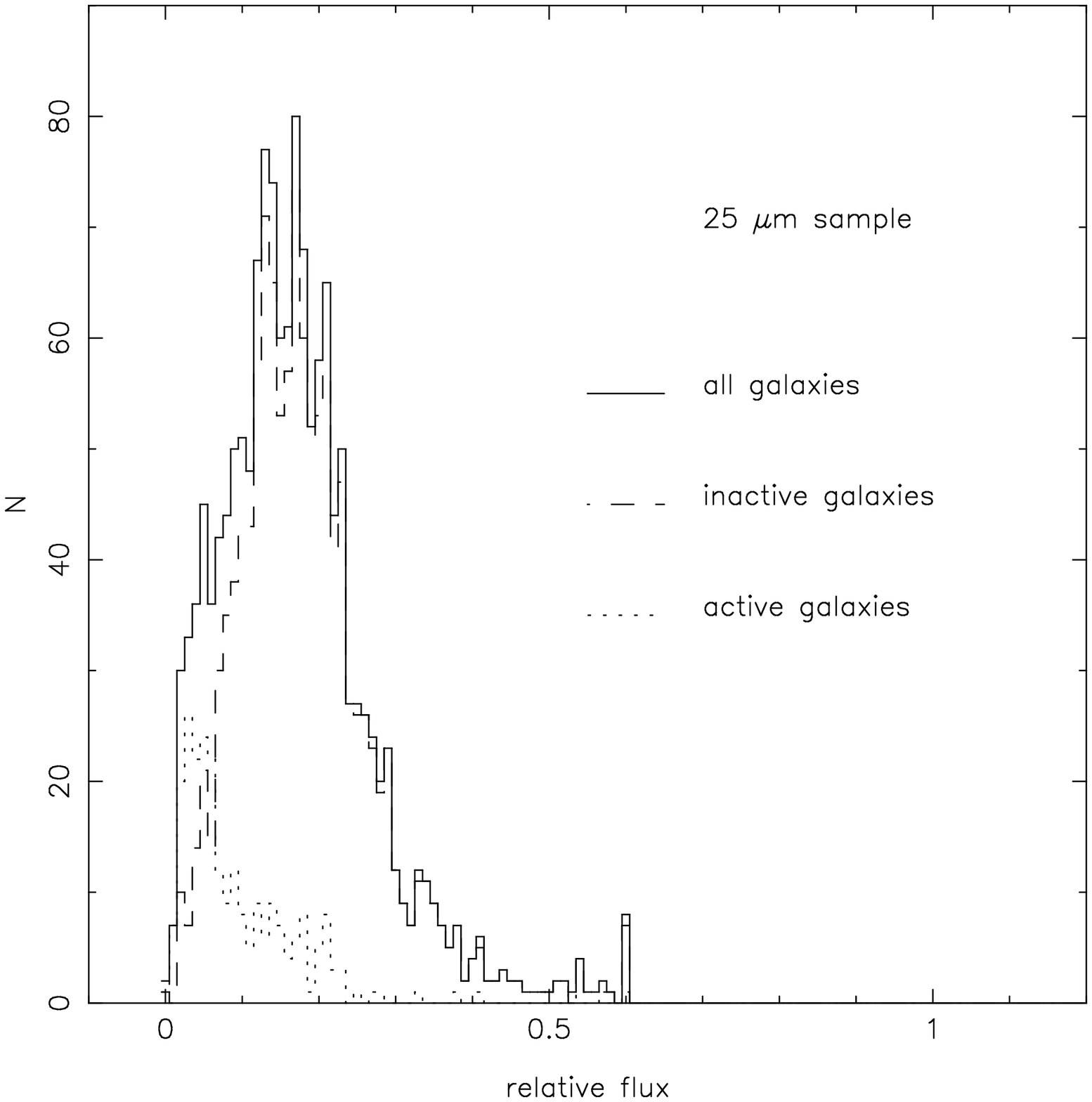}
\caption{
The histogram of the flux due to the (redshifted) emission features
within the IRAS 12 $\mu$m band relative to the 12 $\mu$m flux,
according to the model discussed in Section \ref{sec:pah}.
The result is shown for a sample selected
at 25 $\mu$m, discussed in Paper I (solid histogram), and its subsamples
of active galaxies (dotted histogram) and the rest of the population --
``inactive galaxies'' (dashed histogram).
The overall average relative flux is about 10-20$\%$.
The maximum relative flux is prescribed to be 60$\%$.
\label{fig:pahhist25}}
\end{figure}

The ratios of the fluxes of the four emission features to the IRAS 12 $\mu$m
fluxes of the 16 galaxies are about $0.4$ to $0.6$.
Therefore we would expect less than $60\%$ relative
corrections due to the emission features within the 12 $\mu$m IRAS band
for nearby galaxies.  This is our practical upper-limit for the
relative flux due to the emission features.  Figure \ref{fig:pahhist25}
illustrates the distribution of the relative emission flux in the
12 $\mu$m band for the 25 $\mu$m selected sample containing 1456 galaxies
discussed in Paper I.  We identified (see Section \ref{sec:clrlum})
active galaxies from the others (``inactive galaxies'' in the Figure)
in the sample and show the corrections separately for these populations.
It is seen that the peak correction due to the emission features is somewhat
smaller for active galaxies, indicating stronger radiation intensity
around the grains in these galaxies.  The overall average in-band flux from
emission features is about 10 - 20$\%$ of the 12 $\mu$m flux.
There are only 8 sources ($\sim 0.5\%$ of the sample) lying at the $60\%$
limit, indicating that the model produces modest estimates for the emission
flux for most of the sources.

For $k$-corrections at the IRAS 12 and 25 $\mu$m bands, we assume a power-law
between the 12 and 25$\mu$m continuum fluxes.  The in-band emission
flux (of the redshifted emission features) is first subtracted from the total
in-band flux to estimate the power-law slope.  After $k$-correction,
the rest-frame in-band emission flux is added back to the continuum
to give the corrected total flux.  A more accurate treatment of correcting
for the emission features will be discussed in our forthcoming paper
(\markcite{xb97}Xu et al. 1997), where we will present the SEDs and
our evolutionary models based on the new $k$-correction.

\section{The mid-infrared color-luminosity relation}
\label{sec:clrlum}

There is a significant correlation between the mid-infrared 12-25 $\mu$m color
and the far-infrared luminosity, as also suggested by the 12-25 $\mu$m vs.
60-100 $\mu$m color-color relation
(\markcite{helou86}Helou 1986; \markcite{soi87}Soifer et al. 1987;
\markcite{both89}Bothun, Lonsdale, \& Rice 1989;
\markcite{soi91}Soifer \& Neugebauer 1991).
To explore a similar color vs. mid-infrared luminosity relation, we use
the large 25 $\mu$m selected samples discussed in Paper I.  We also
incorporate the model discussed previously to calculate the
fluxes and luminosities more accurately.
The luminosities are calculated similarly as we did in Paper I.

A 25 $\mu$m selected all-sky sample may suffer from incompleteness due to
large noises in 25 $\mu$m near the ecliptic plane.  We have demonstrated
in Paper I that our 1456-galaxy sample is highly complete across the
ecliptic plane.  We have also defined, in Paper I, a high-quality sample
which has a higher 400 mJy flux density limit near the ecliptic region
(the rest of the region is limited at 250 mJy).  Here we want to compare
the results of the color-luminosity relation for these different samples,
and then we continue the analyses using the high-quality sample, which
contains 1049 galaxies.

Since the samples are selected with moderate or high quality detections
at 25 $\mu$m, some galaxies have upper limits for flux densities
at 12 $\mu$m.  Therefore the mid-infrared colors of these galaxies
are so-called censored data (\markcite{feig85}Feigelson \& Nelson 1985;
\markcite{schmt85}Schmitt 1985; \markcite{isob86}Isobe et al. 1986).
We use the standard Kaplan-Meier (K-M) estimator 
(\markcite{kap58}Kaplan \& Meier 1958) to calculate the average color
at each (binned) luminosity, assuming that the censored data has a random
distribution at that luminosity.  The Buckley-James (B-J) linear regression
(\markcite{buck79}Buckley \& James 1979) is also used to provide a
linear interpretation of the color-luminosity relation. It is calculated
from the original data, not from the K-M estimates.
We have used the survival analysis tools in the IRAF
STSDAS/ANALYSIS/STATISTICS package to carry out the analysis.

\begin{figure}
\plotone{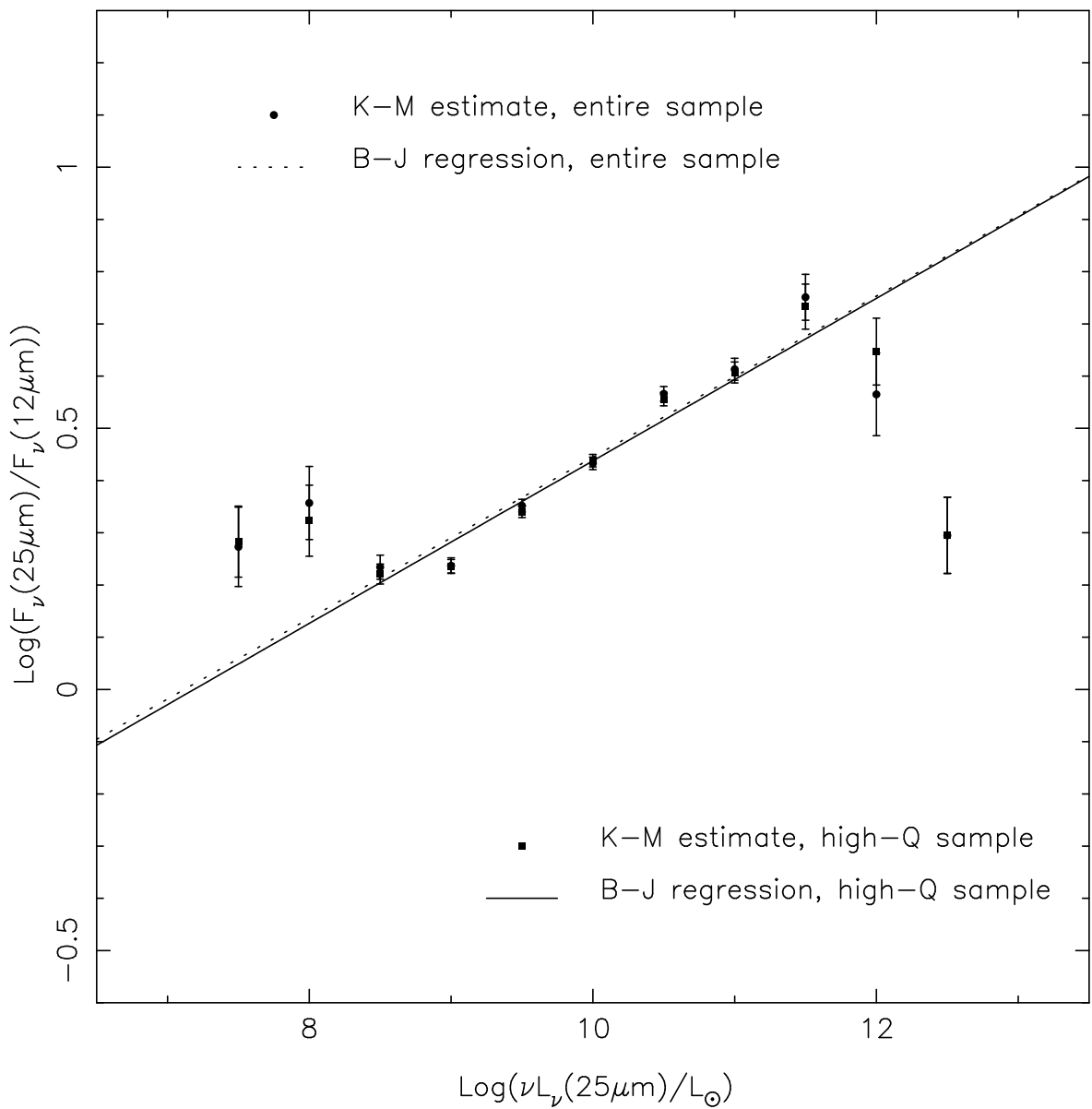}
\caption{
The 12-25 $\mu$m color vs. 25 $\mu$m luminosity relation for the
entire 25 $\mu$m limited sample and the high-quality sample discussed
in Paper I.  The heavy dots show the Kaplan-Meier estimates of the mean
colors in the half-decade luminosity bins for the entire sample, and squares
for the high-quality sample.  The solid and dotted lines are the
Buckley-James linear regressions of the color-luminosity relation
for the high-quality and the entire samples, respectively.
The results are nearly identical for the two samples.
\label{fig:km25pah}}
\end{figure}

We first study the relation between the 12-25 $\mu$m color and the
25 $\mu$m luminosity.
Figure \ref{fig:km25pah} compares the results
for the high-quality sample and the entire 25 $\mu$m sample limited
at 250 mJy.  Here the monochromatic luminosities are expressed
as $\nu L_\nu$ and have units of solar luminosities.  The squares
are the K-M estimates in each half-decade luminosity bin.  There is
clearly a correlation between the 12-25 $\mu$m color and the 25 $\mu$m
luminosity.  The results
for the two samples are similar.  The B-J linear regression gives
indistinguishable slopes and intercepts for the linear interpretation.
The B-J results for these and the
subsequent analyses are listed in Table \ref{tab:bj}.
This confirms a conclusion from Paper I that the 250 mJy-limited sample,
while at lower signal-to-noise ratio than the high-quality sample,
does not contain systematic flux or color errors.

\begin{figure}
\plotone{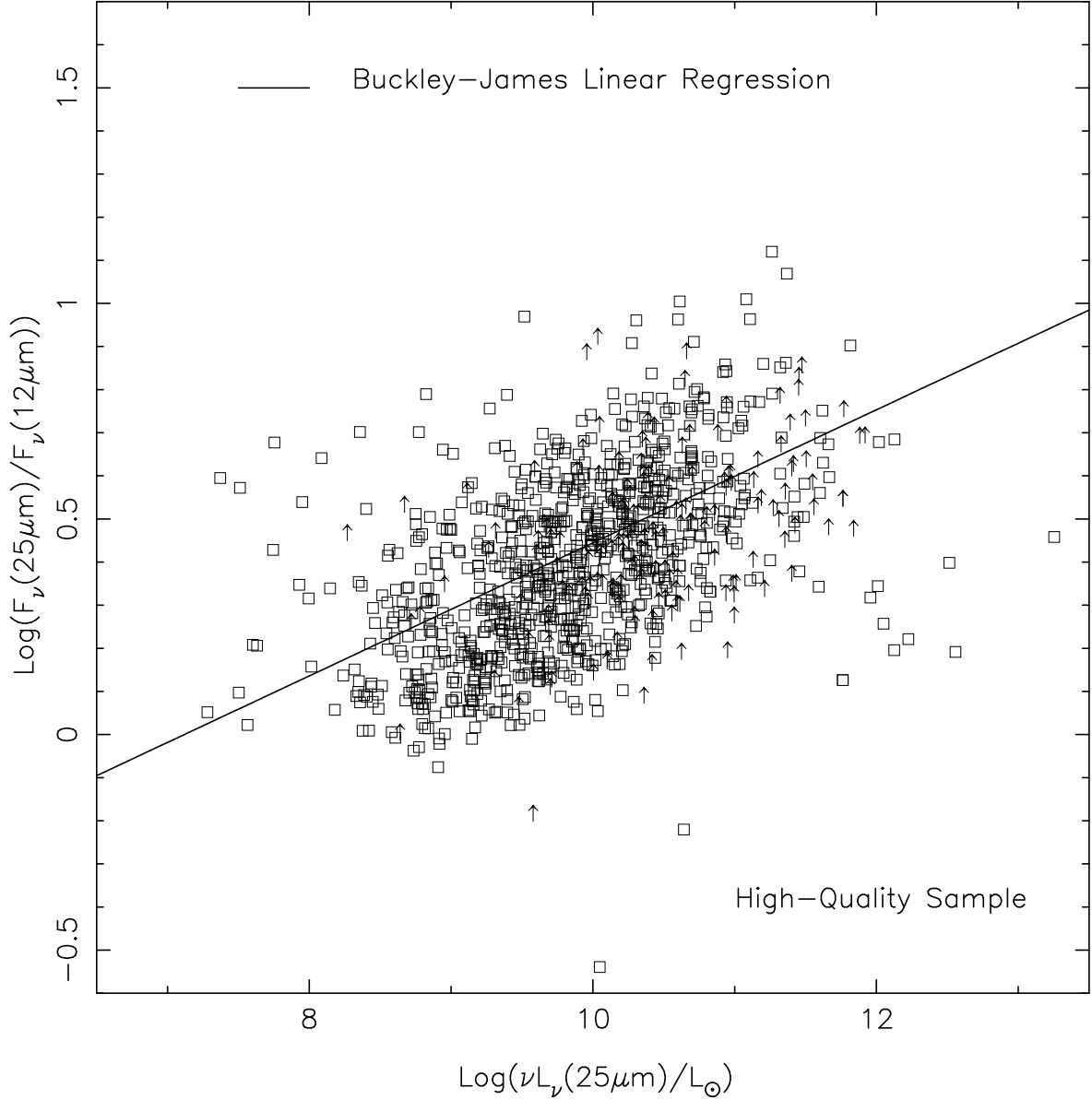}
\caption{
The scattered diagram for the 12-25 $\mu$m color vs. 25 $\mu$m luminosity
relation obtained using the high-quality 25 $\mu$m selected sample.
Arrows indicate the upper-limit detections at 12 $\mu$m.  The solid line
is the Buckley-James linear regression, the same as the dashed line in
Figure \ref{fig:km25pah}.
\label{fig:scatHQ25pah}}
\end{figure}

\begin{figure}
\plotone{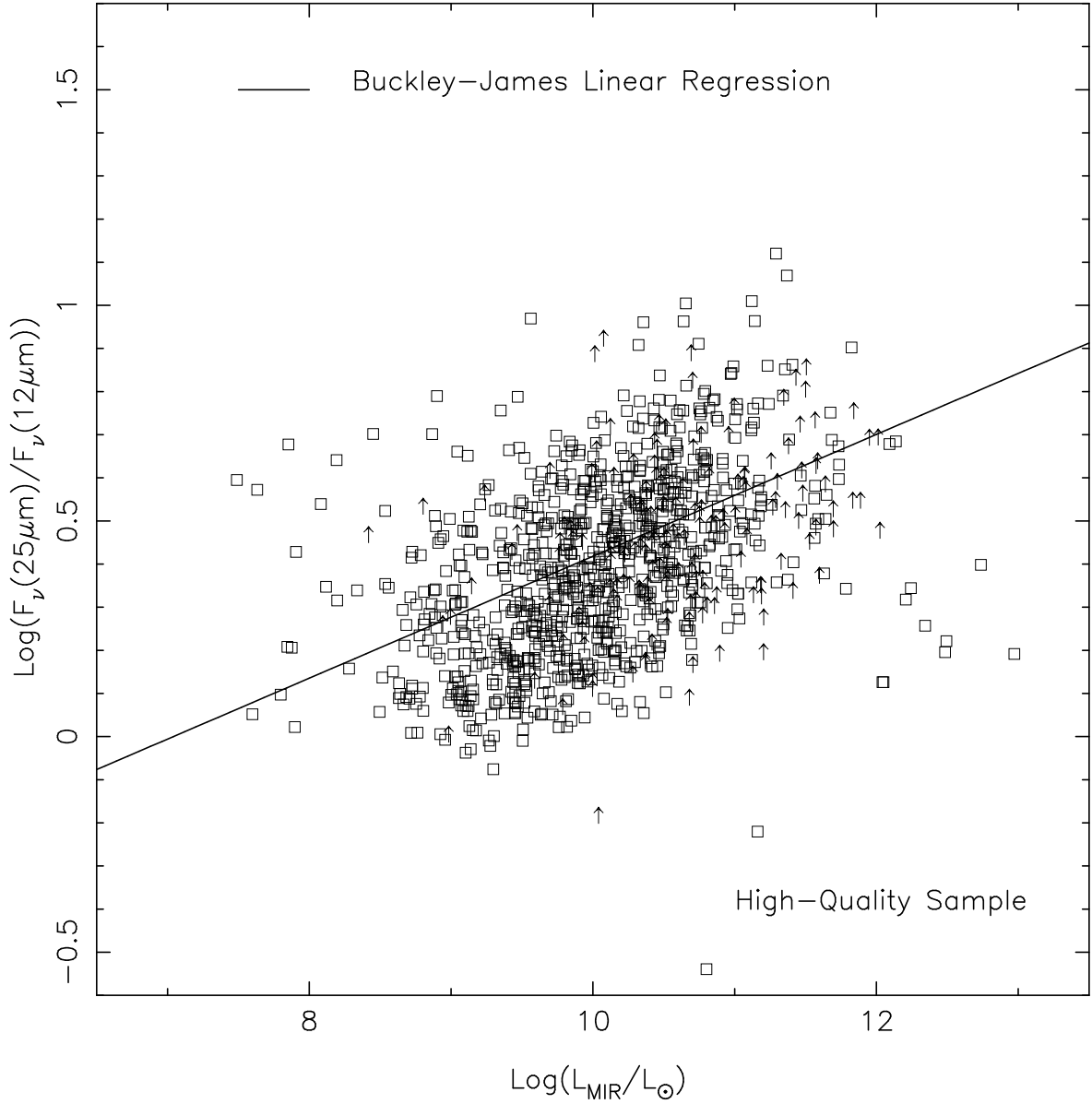}
\caption{
The 12-25 $\mu$m color vs. the mid-infrared luminosity relation obtained
for the 25 $\mu$m high-quality sample.  The mid-infrared flux is defined
in the text.  The scattering patterns are similar to those in Figure
\ref{fig:scatHQ25pah}.
\label{fig:scatHQmirpah}}
\end{figure}

Figure \ref{fig:scatHQ25pah} shows the scattered color-luminosity
data and the B-J linear regression for the high-quality sample.
To examine whether there are any artificial factors in the relation
caused by the proportionality of the 25 $\mu$m fluxes and luminosities,
we also obtained the relation of the same color vs. 60 $\mu$m luminosity
for the sample.  The results are shown in Table \ref{tab:bj}.  It turns out
that the B-J slope hardly changes, while the intercept decreases,
indicating a roughly constant 25 to 60 $\mu$m ratio as a function
of luminosity, which was discovered earlier (Soifer \& Neugebauer 1991).
This confirms the reliability of the color vs.
25 $\mu$m luminosity relation.  Furthermore, we have calculated the
mid-infrared fluxes of the galaxies in the high-quality sample,
according to the following definition, given by
\markcite{xb95}Xu \& Buat (1995):
\begin{equation}
\label{eqn:mir}
F_{mir}=F_{12\mu m}\frac{c\delta\lambda_{12\mu m}}{\lambda_{12\mu m}^{2}}+
F_{25\mu m}\frac{c\delta\lambda_{25\mu m}}{\lambda_{25\mu m}^{2}},
\end{equation}
where $F_{mir}$, $F_{12\mu m}$, $F_{25\mu m}$ are mid-infrared flux,
12 $\mu$m and 25 $\mu$m flux densities, respectively, $\lambda$s are
wavelengths, c is the speed of light, $\delta\lambda_{12\mu m}=7\mu m$,
and $\delta\lambda_{25\mu m}=25 \mu m$ are the band passes.
$\delta\lambda_{25\mu m}$ is increased from the real IRAS band pass to
cover the spectral range up to 40 $\mu$m.  The resulting MIR
color-luminosity relation is shown in Figure \ref{fig:scatHQmirpah}.
The similar proportionality between the 12-25 $\mu$m color and the MIR
luminosity and the scattering patterns exist, comparing with those in
Figure \ref{fig:scatHQ25pah}.  But this relation is more robust
to error in the flux measurement in one of the MIR bands.
The B-J regression results are listed in Table \ref{tab:bj}.

\begin{figure}
\plotone{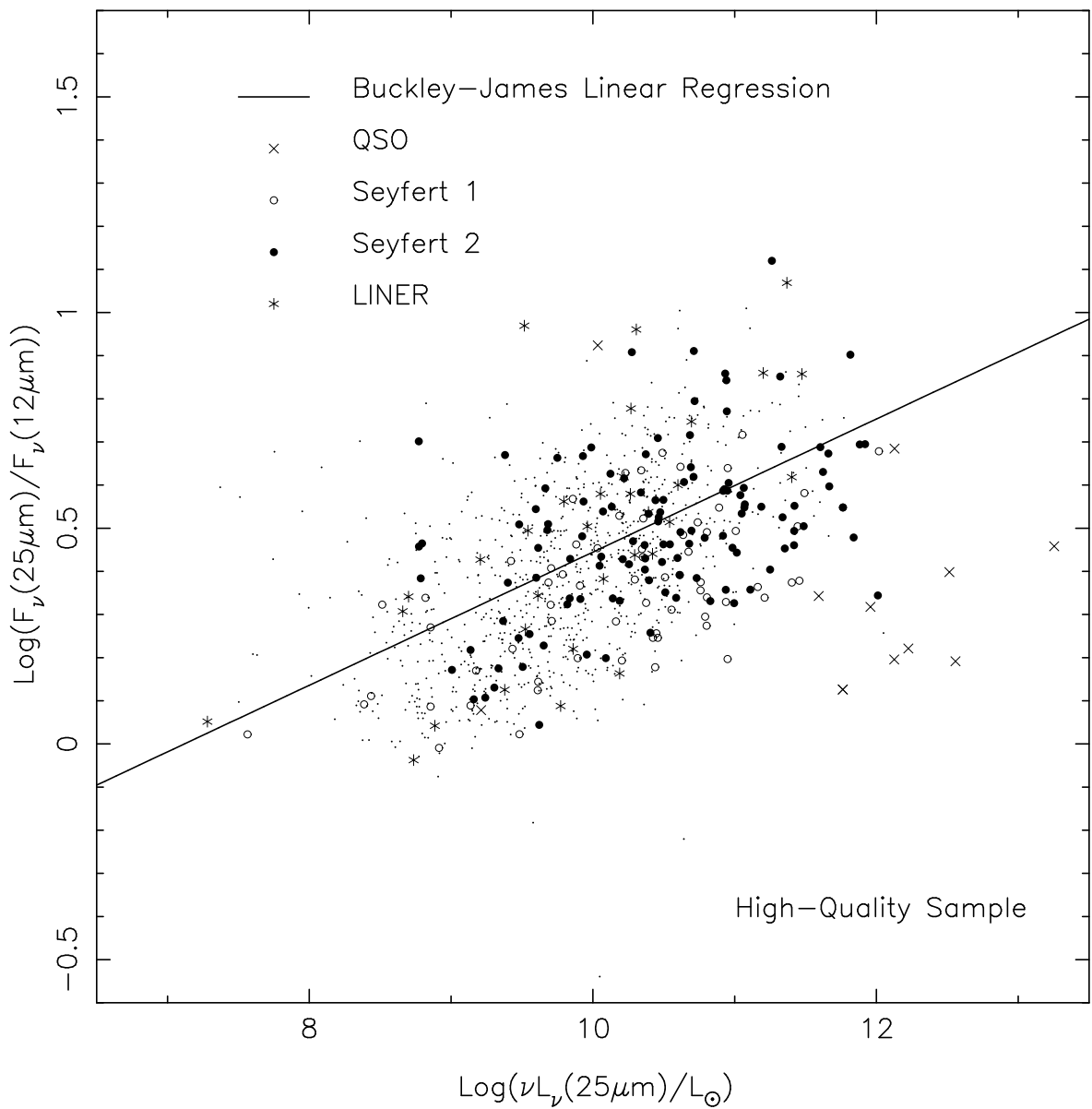}
\caption{
The populations of Seyfert galaxies (circles for Seyfert 1s, heavy dots
for Seyfert 2s), QSOs (crosses), and LINERs (stars) in the same
color-luminosity relation as in Figure \ref{fig:scatHQ25pah}.
All the other galaxies are indicated by light dots.
A population of infrared-luminous quasars has systematically bluer
color and groups together below the Buckley-James linear regression line.
\label{fig:clrlumHQallpop}}
\end{figure}

To study the distribution of different galaxian populations
in the color-luminosity relation, we have obtained the morphology and
type information for the sources in the high-quality sample from the
NASA/IPAC Extragalactic Database (NED).
In Figure \ref{fig:clrlumHQallpop} we re-draw the same color-luminosity
relation as in Figure \ref{fig:scatHQ25pah}, with the populations of
QSOs, Seyfert 1 and 2s, and LINERs highlighted.  The Seyfert 2s and LINERs
appear to follow the B-J regression, but Seyfert 1s tend to fall below the
regression line on average.  There is a population of luminous quasars grouped
below the regression line (very blue), indicating that they have
relatively flat spectra.  This causes the drop of the color values
of the two highest luminosity bins in Figure \ref{fig:km25pah}.
On the other hand, the nearby low luminosity infrared sources
are redder, causing the deviation of the color-luminosity relation
in the first few bins in Figure \ref{fig:km25pah}.

\begin{figure}
\plotone{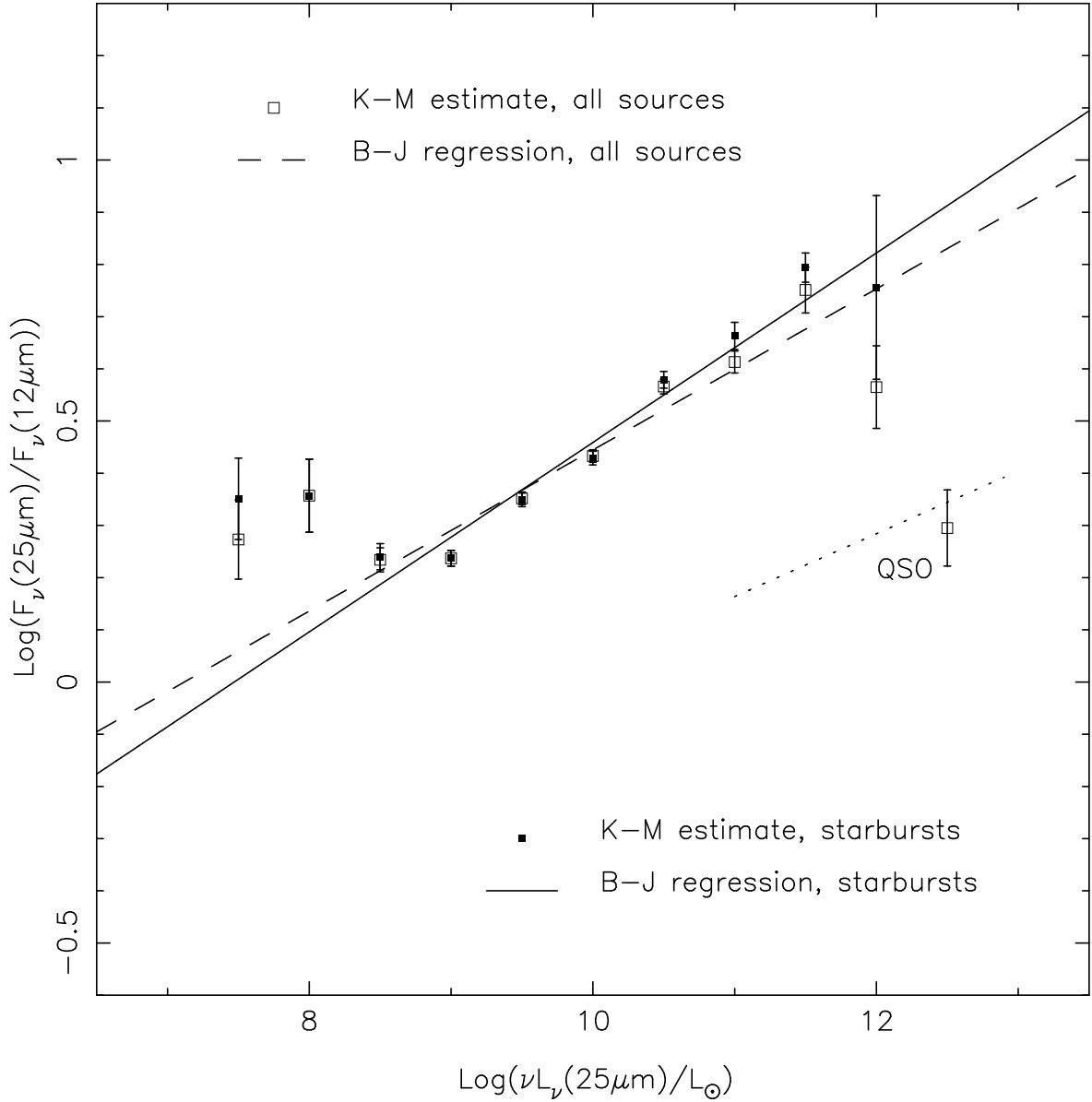}
\caption{
The Kaplan-Meier estimate and the Buckley-James linear regression
of the color vs. 25 $\mu$m luminosity relation.  Empty squares and dashed
lines show the result for the whole population and filled squares and solid
lines for the starburst galaxy population.  The dotted line is the linear
regression for the luminous quasars.
\label{fig:km25comp}}
\end{figure}

To further investigate the effects of different populations on
the color-luminosity relation, we exclude the Seyfert galaxies, LINERs,
and quasars from the sample and calculate the K-M average and B-J regression
for the rest of the galaxies, which are essentially starburst galaxies
and luminous infrared galaxies
(defined as $L_{fir} \stackrel{>}{_{\sim}} 10^{11}L_{\odot}$).
Figure \ref{fig:km25comp} compares the results with those obtained
for all populations.  The linear regression for the infrared luminous
quasars is also shown by the dotted line. The average slope of the
color-luminosity relation becomes greater for the starburst population. 
At high-luminosity bins, the average colors of the starburst galaxies
become systematically redder, and much redder than those of the quasars.
We notice that there is a color turn-over at the $10^{12}L_{\odot}$ bin,
but statistical significance is poor with only two sources in that bin.
Higher luminosity bins are completely occupied by quasars in our sample.
This relation implies that very high luminosity starburst systems, if
they exist, could be distinguished from quasars by
their 12-25 $\mu$m color.

\begin{figure}
\plotone{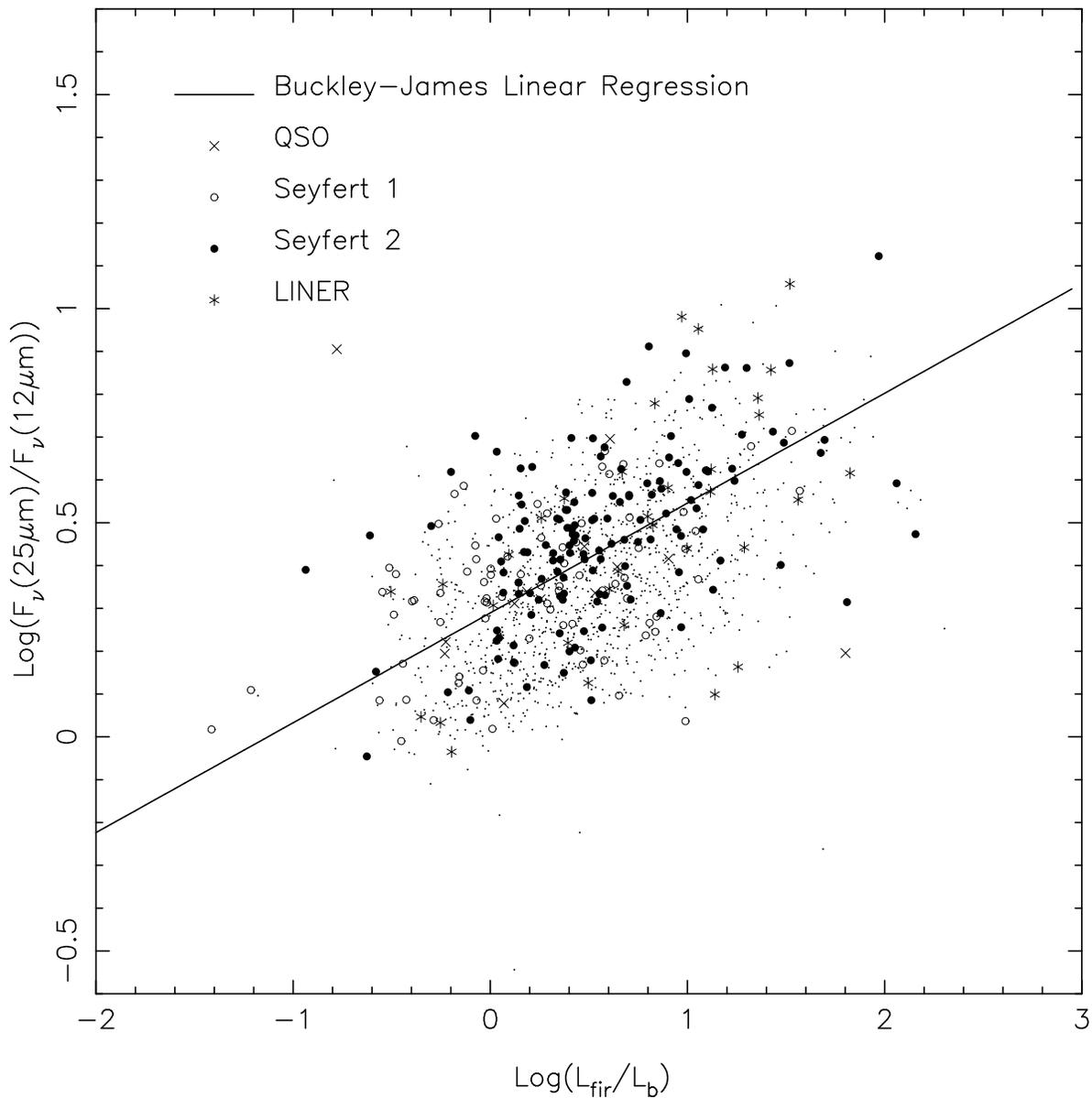}
\caption{
The scattered diagram for the relation between the 12-25 $\mu$m color
vs. the ratio of far-infrared and the blue luminosities, with the same
populations in Figure \ref{fig:clrlumHQallpop} indicated.
The quasars identified in Figure \ref{fig:clrlumHQallpop} now
distribute around the linear regression line,
indicating their greater blue luminosities.
\label{fig:firfbpop}}
\end{figure}

\begin{figure}
\plotone{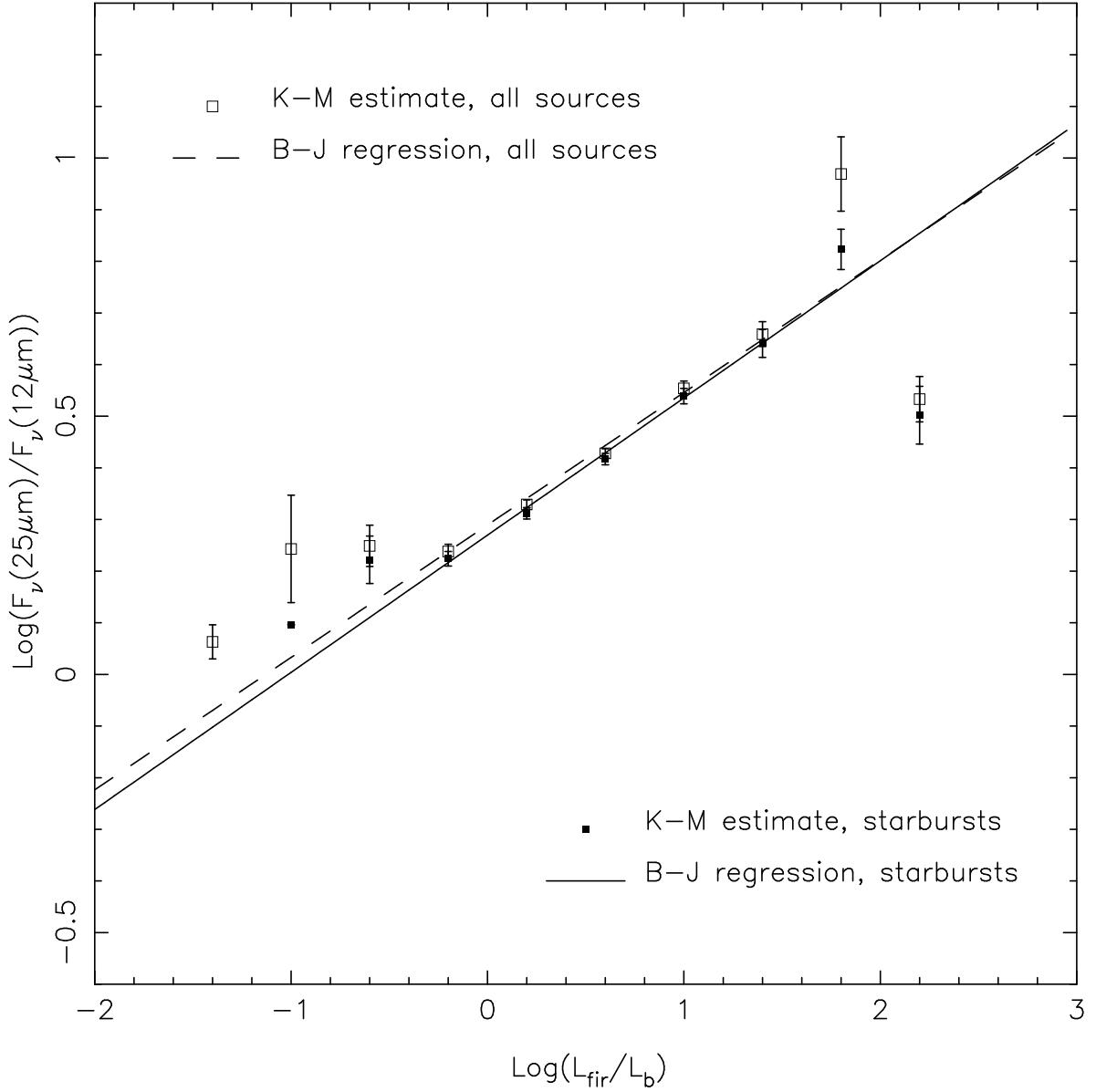}
\caption{
The Kaplan-Meier estimate of the average 12-25 $\mu$m color vs. the
ratio of far-infrared and blue luminosities.
The squares show the K-M average colors at the given
$L_{fir}/L_{b}$ bins. The empty squares are calculated for all sources,
and the filled squares for starburst galaxies only.
The dashed and solid line show the Buckley-James
linear regression for all sources and for starbursts, respectively.
The blue magnitudes are obtained for 1390
galaxies in the 25 $\mu$m-limited sample, as described in the text.
\label{fig:fir2km}}
\end{figure}

We have also obtained the blue magnitudes
for 1390 galaxies in our entire 250 mJy-limited sample.
These magnitudes were obtained from the following catalogs in order of
preference: the CfA redshift catalog, the COSCAT, and the NED.
We then calculated the relation between the 12-25 $\mu$m color and the
ratio of the far-infrared and the blue luminosity, which indicates
the starburst strength.  Figure \ref{fig:firfbpop} shows the B-J
linear regression with all populations included.  The active galaxies
and quasars are indicated similarly as in Figure \ref{fig:clrlumHQallpop}.
The figure shows that a good correlation exists between these colors.
Different populations, including Seyfert 1 galaxies and the infrared
luminous quasars, appear to follow the same linear regression.
A single 12-25 $\mu$m color vs. $L_{fir}/L_{b}$ relation appears to
apply equally to the different populations.
To examine this, we have re-calculated the K-M average and B-J regression
for the population of starburst galaxies only (excluding Seyfert galaxies,
quasars, and LINERs).  The results are shown in
Figure \ref{fig:fir2km}.  The 12-25 $\mu$m color-$L_{fir}/L_{b}$ relations
for the starbursts and all populations are essentially the same.
As mentioned in Section \ref{sec:pah}, starbursts and activities in the
galactic nuclei can have a similar effect on the grain emission.  The
$L_{fir}/L_{b}$ is a good strength indicator (see below) for the starburst
population, whereas the color vs. 25 $\mu$m luminosity is more sensitive
to these different populations.
We also obtained the 12-25 $\mu$m color vs. $L_{mir}/L_{b}$ relation
using the definition of the mid-infrared flux in Equation \ref{eqn:mir},
and the results are similar to Figure \ref{fig:firfbpop}.

\begin{figure}
\plotone{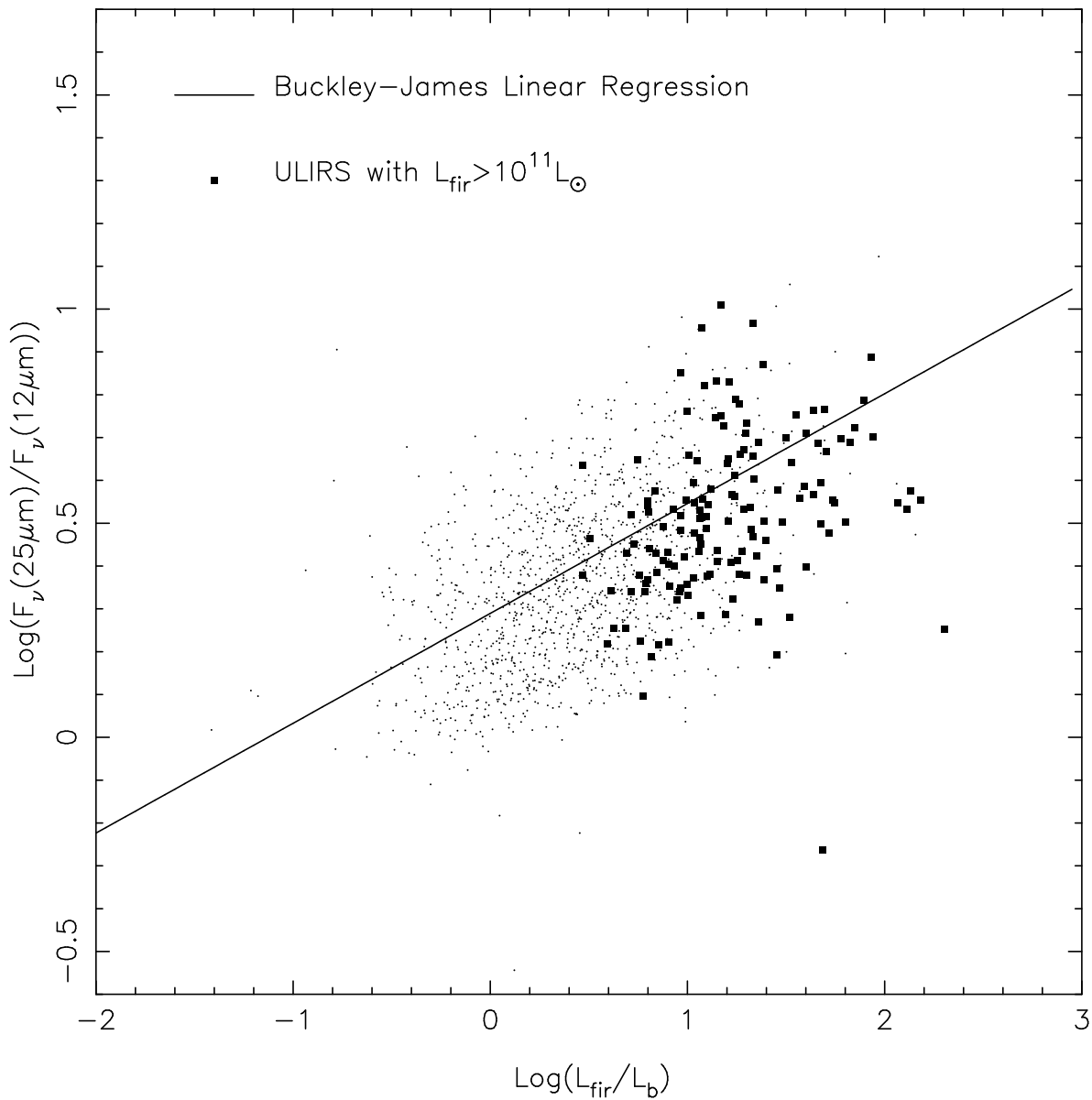}
\caption{
The same color-luminosity relation as in Figure \ref{fig:firfbpop}
with the luminous and ultraluminous infrared galaxy population (ULIRS,
$L_{fir}\ge 10^{11}L_{\odot}$) identified (filled
squares).  The rest of the sources are indicated by light dots.
It is clearly seen that the luminous and ultraluminous infrared galaxies
occupy the region with high $L_{fir}/L_{b}$ ratios.
\label{fig:ulirs}}
\end{figure}

The 12-25 $\mu$m color vs. 25 $\mu$m luminosity relation in
Figure \ref{fig:km25comp} provides powerful tools in recognizing different
populations such as starbursts, luminous and ultraluminous infrared galaxies,
and quasars.  Infrared luminous quasars can be easily identified by
their bluer colors in the diagram.  The luminous and ultraluminous infrared
galaxies would generally have redder colors at high-luminosity bins
in such a relation.  Although different populations
are well-mixed in the color-color relation of Figure \ref{fig:fir2km},
the ultraluminous infrared galaxies also segregate themselves in such a
relation, as suggested by the proportional relation between the ratio
$L_{fir}/L_{b}$ and $L_{fir}$, found by Soifer et al. (1987).
Figure \ref{fig:ulirs} demonstrates such a segregation.  In the Figure, we have
identified the luminous and ultraluminous infrared galaxies
(excluding active galaxies and quasars) with $L_{fir}\ge 10^{11}L_{\odot}$
(filled squares) among the rest of the population.  It is seen that all
galaxies in this population have greater $L_{fir}/L_{b}$ ratios.
They can be selected from high $L_{fir}/L_{b}$ bins with high probability.
For example, among 224 sources with log$_{10}(L_{fir}/L_{b}) > 1$ in the
Figure, 133 sources have $L_{fir}\ge 10^{11}L_{\odot}$, and 100 of them
are luminous and ultraluminous infrared galaxies.  The ease of detecting such
luminous star-forming galaxies is what makes modest-sized infrared telescopes
such effective probes of galaxy evolution in the early universe, compared to
the strong selection effects that complicate optical and UV studies.

The Seyfert galaxies, on the other hand, are more difficult to identify,
although the difference in the linear regressions in Figure \ref{fig:km25comp}
is partly due to this population (Seyfert 1s are mostly below the regression
lines). The luminous Seyfert galaxies are also more scattered in the
12-25 $\mu$m color vs. $L_{fir}/L_{b}$ ratio relation than the ultraluminous
infrared galaxies, increasing the probability of identifing the
latter at high $L_{fir}/L_{b}$ ratio bins.

\section{The local luminosity function at 12$\mu$m}
\label{sec:lumfcn}

\subsection{Sample Selection}
\label{sec:sample}

We based our sample on a selection from the IRAS Faint Source Survey
(FSS; \markcite{mos92}Moshir et al. 1992).  The main data product of the
FSS is the Faint Source Catalog (FSC) and the Faint Source Reject File.
The Faint Source Reject File contains possible detections that were not
included in the FSC for assorted quality-control reasons.

Subsamples can be drawn from the FSS-based sample flux-limited at 250 mJy
at 25 $\mu$m discussed in Paper I.  A 12 $\mu$m flux-limited
subsample drawn directly from that sample, however, would suffer from
two biases.  First, since we used moderate or good quality detection
criterion in the 25 $\mu$m sample, galaxies with moderate or good quality
detections at 12 $\mu$m may not have the same or better detection qualities at
25 $\mu$m, and therefore would not be included in such a subsample.  The
second source of incompleteness for such a subsample comes from a color
bias: sources which satisfy the 12 $\mu$m flux-density criterion may not have
25 $\mu$m flux densities greater than 250 mJy, thereby would be excluded.
On the other hand, we want to stay close to the spatial and color selection
criteria which defined the highly complete 25 $\mu$m sample
(see Paper I for discussions of the completeness of the sample).
We therefore used the following criteria:

\[ \begin{array}{l}
    F_{\nu}(12) \ge 200  {\rm mJy;} \\
    \mbox{moderate or good quality detection at}~12\mu{\rm m;} \\
    |b| \ge 30^\circ, \mbox{and not in \markcite{str90}Strauss et al. (1990) \ excluded zone}; \\
    F_{\nu}(25) < F_{\nu}(60)~{\rm and}~F_{\nu}(12) < 2F_{\nu}(25), ~ {\rm or}~  \\
    F_{\nu}(60) < F_{\nu}(25) < 1.6F_{\nu}(60)~{\rm and} ~F_{\nu}(12) < F_{\nu}(25).
\end{array} \]

The $90\%$ completeness limit for the FSC lies at a 12 $\mu$m flux density
of 180 mJy for sky covered by 2 HCONs, and at 150 mJy for coverage of
3 HCONs, for $|b| > 10^{\circ}$ (Moshir et al. 1992).  Here the flux-density
limit of 200 mJy is chosen because it gives the sample a high
redshift-completeness (see below).

1059 sources in the FSC meet the above criteria.  We found 619 galaxies
which are in the 25 $\mu$m sample from Paper I with redshifts.  We matched the rest
of the sources with the 1.2 Jy Survey catalog, the November
1993 public version of J.P. Huchra's ZCAT, the NED database, and the
SIMBAD database.  This process resulted in an addition of 43 galaxies with
redshifts, 3 galaxies with no redshift information, and the identification
of the rest of the sources as non-galaxies (mostly stars).

We have applied our selection criteria to the Faint Source
Reject File and found 101 sources.  We followed the same matching techniques
as we used in Paper I to find galaxies from the Reject File and
found 6 galaxies, all with redshifts.  A sample of 101 sources randomly
distributed in the same sampling region would result in less than one match
with galaxies.  This gives a total of 668 galaxies with redshifts in our
final sample, and the redshift completeness of the sample is $99.5\%$.

The 1.2 Jy Survey catalog also provides the ADDSCAN
flux densities for the sources extended at $60\mu\rm m$.
For those sources which have no ADDSCAN flux densities in the 1.2 Jy Survey
but are extended at $12\mu\rm m$, we obtained their ADDSCAN flux densities
by running XSCANPI. These ADDSCAN flux densities are used in the calculation
of the 12 $\mu$m luminosity function.

\begin{figure}
\plotone{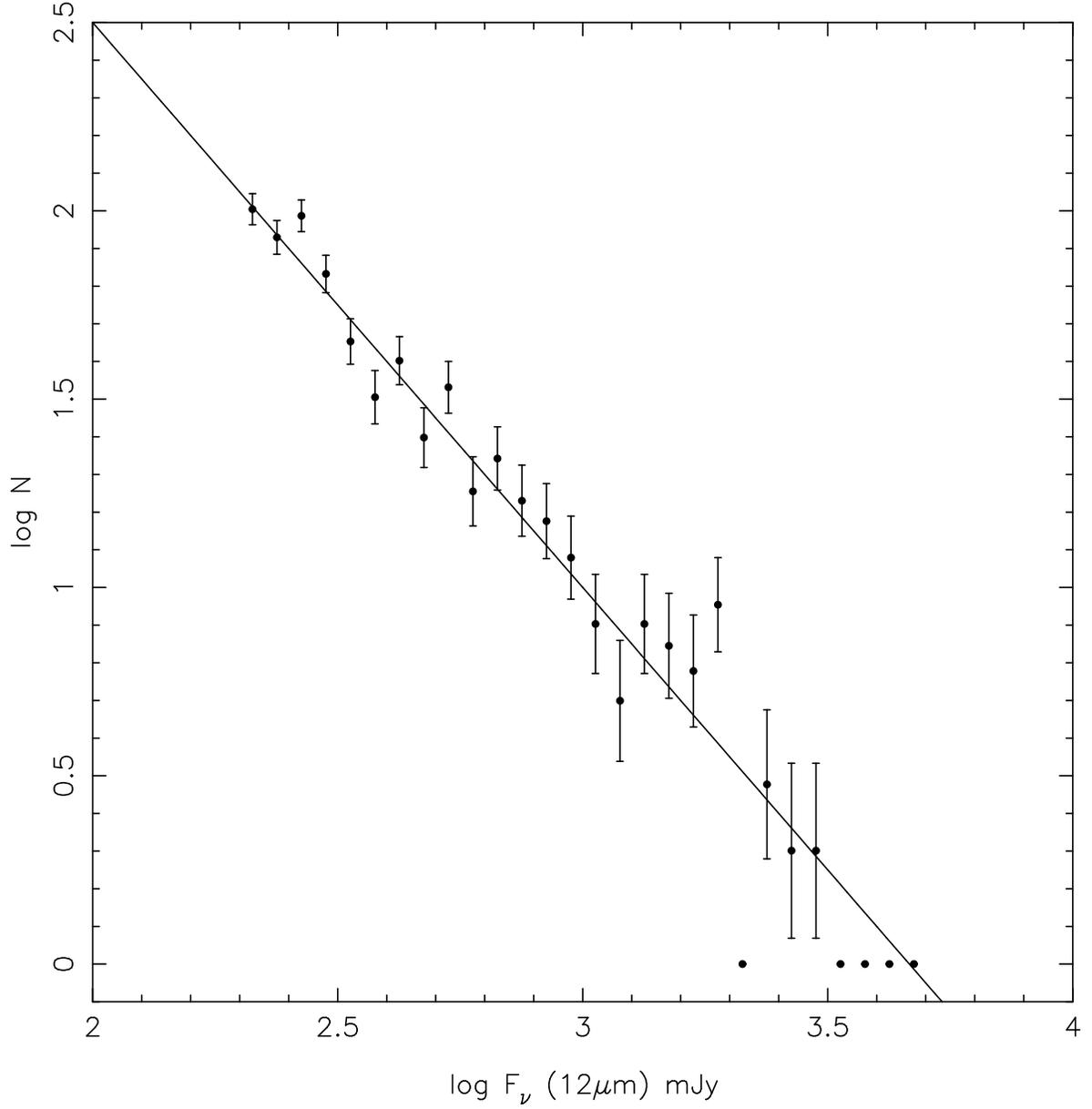}
\caption{
The number-flux test for the 12 $\mu$m selected sample flux-density limited
at 200 mJy.  For a complete, spatially homogeneously distributed
sample, the relation should have a slope of -1.5, the solid line.
The sample closely follows this relation.
\label{fig:comp}}
\end{figure}

In Figure \ref{fig:comp} we illustrate a number vs. flux test to examine
the completeness of our sample.  It is seen that the source counts
in different flux-density bins fall close to the line with the slope -1.5,
expected for a homogeneous flux-limited sample of galaxies.  There is no
indication of incompleteness of the sample down to 200 mJy.

The model discussed in Section \ref{sec:pah} is incorporated
to estimate the $k$-corrections.  The relative flux due to
the emission features between 3 $\mu$m and 11.6 $\mu$m is peaked
at $\sim 18\%$, similar to the features in Figure \ref{fig:pahhist25}.

\subsection{Luminosity function results}
\label{sec:pahlf}

To calculate the luminosity function for the sample,
we have used secondary distances when available.
Otherwise, a linear Virgocentric inflow model
(\markcite{aar82}Aaronson et al.\ 1982) and a Hubble
constant of 75 km~s$^{-1}$ Mpc$^{-1}$ is used. (Based on the same reasons
we outlined for the 25 $\mu$m selected sample in the Appendix of Paper I,
we note that our estimated luminosities here can be simply scaled
if a different Hubble constant is used.)
We have followed the same criteria as \markcite{soi87}Soifer et al. (1987)
to identify Virgo cluster galaxies (with a distance of 17.6 Mpc).

We have calculated the 12 $\mu$m luminosity function of the sample
using the $1/V_{max}$ estimator, the non-parametric maximum-likelihood
method, and the parametric maximum-likelihood method with the
Yahil et al.\ \markcite{yah91} (1991) robust parametric form.
As discussed in Paper I, the $1/V_{max}$ estimator is potentially
affected by the inhomogeneities of the large-scale spatial
distribution of the sources in the sample.  The maximum-likelihood
methods assume that the spatial and luminosity distributions are independent,
so that the shape of the luminosity function can be determined without
being affected by any spatial structures, whereas the normalization
of the function is lost.  In this paper, we normalize these luminosity functions
to the total number of sources in our sample.

\begin{figure}
\plotone{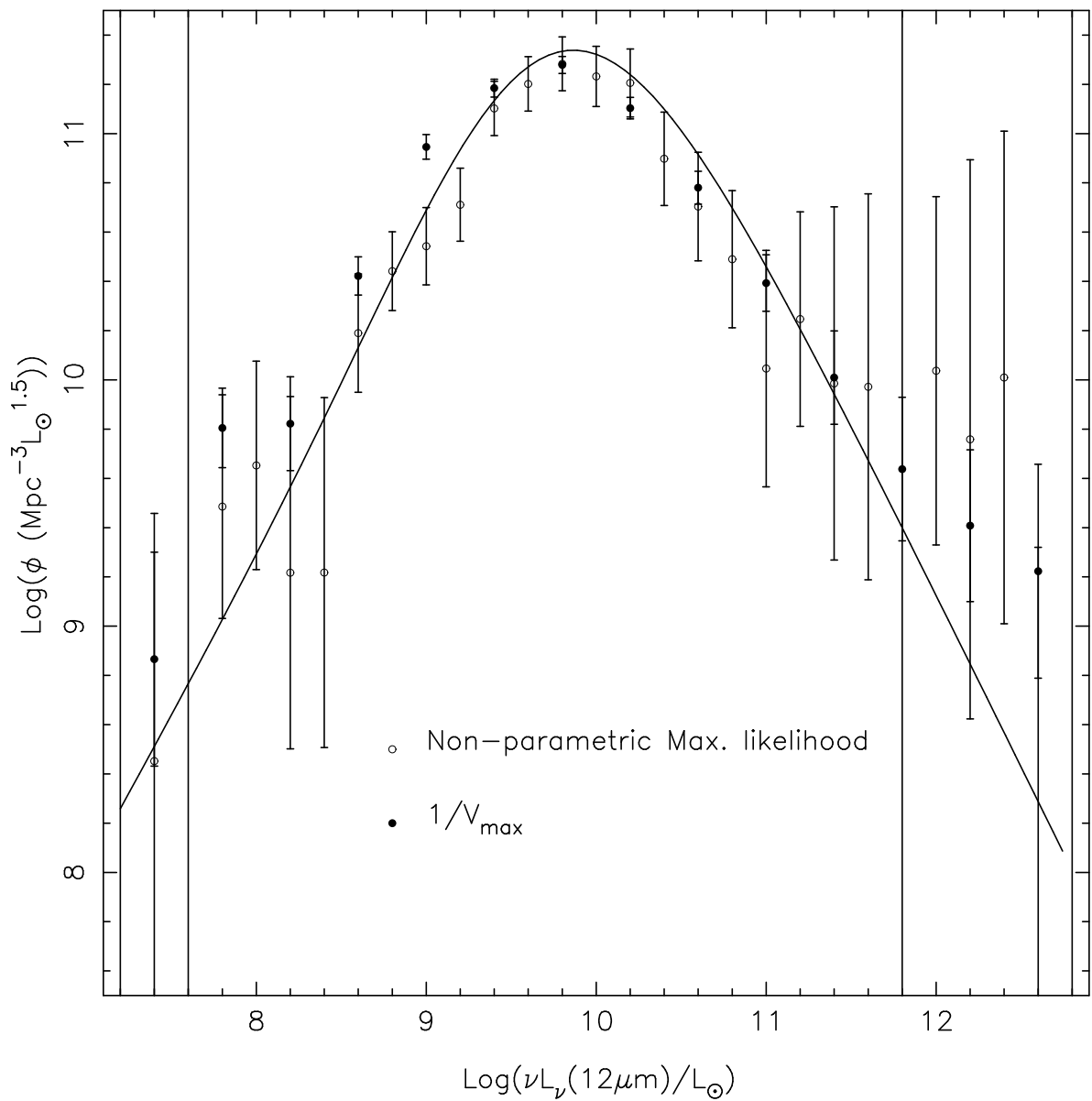}
\caption{
The 12 $\mu$m visibility function calculated using the $1/V_{max}$
estimator (heavy dots), the parametric (solid line) and the non-parametric
(circles) maximum-likelihood methods.  The parametric form of Yahil et al.
(1991) is used.  It does not fit the flattened high-L end of the
other estimates.  The maximum-likelihood estimates are normalized to
the total number of sources in the sample.  The 1-$\sigma$ error bars
are shown.
\label{fig:lfrawpah}}
\end{figure}

Figure \ref{fig:lfrawpah} shows the results.  The luminosity function
is plotted in the form of a visibility function (luminosity function
multiplied by a factor of $L^{2.5}$),
which has the advantage of showing the relative numbers of galaxies
in a flux-limited sample.  The results of the $1/V_{max}$ estimator
agree with those of the non-parametric maximum-likehood method within
the 1-$\sigma$ errors, although at high luminosity bins the non-parametric
maximum-likelihood method gives higher values at the first few bins
then drops off, and the error bars become very large in these bins.
The large uncertainty in these bins is one disadvantage of the method
when dealing with small numbers of sources.
The results of the parametric maximum-likelihood method,
however, show a steep slope after the peak and do not agree at
higher luminosity bins. (The parameters are: $C=0.00252$, $\alpha=0.2648$,
$\beta=2.6372$, $L_{\star}=4.9215\times 10^{9}L_{\odot}$, using the same
definitions as in Yahil et al. 1991 and Paper I) This indicates that
this parametric form may not apply to the 12 $\mu$m luminosity function.

\begin{figure}
\plotone{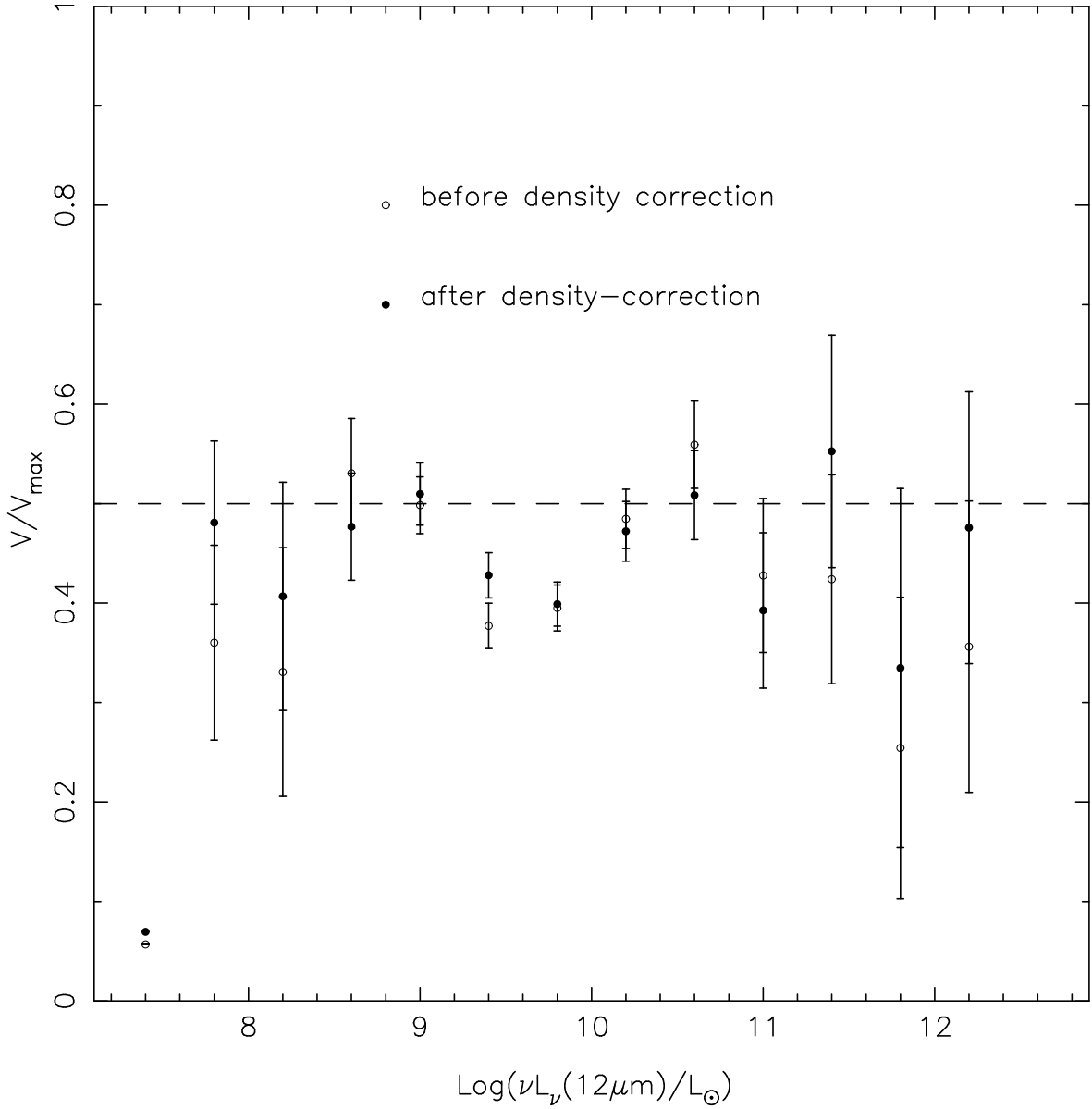}
\caption{
The $V/V_{max}$ test for the 12 $\mu$m sample.  Results are shown for
both before (circles) and after (heavy dots) the corrections for
the radial density inhomogeneities.  The values of $V/V_{max}$ are
closer to 0.5 after the density correction, which indicates the better
quality of the $1/V_{max}$ estimates in Figure \ref{fig:lfpah}.
\label{fig:vovmax}}
\end{figure}

\begin{figure}
\plotone{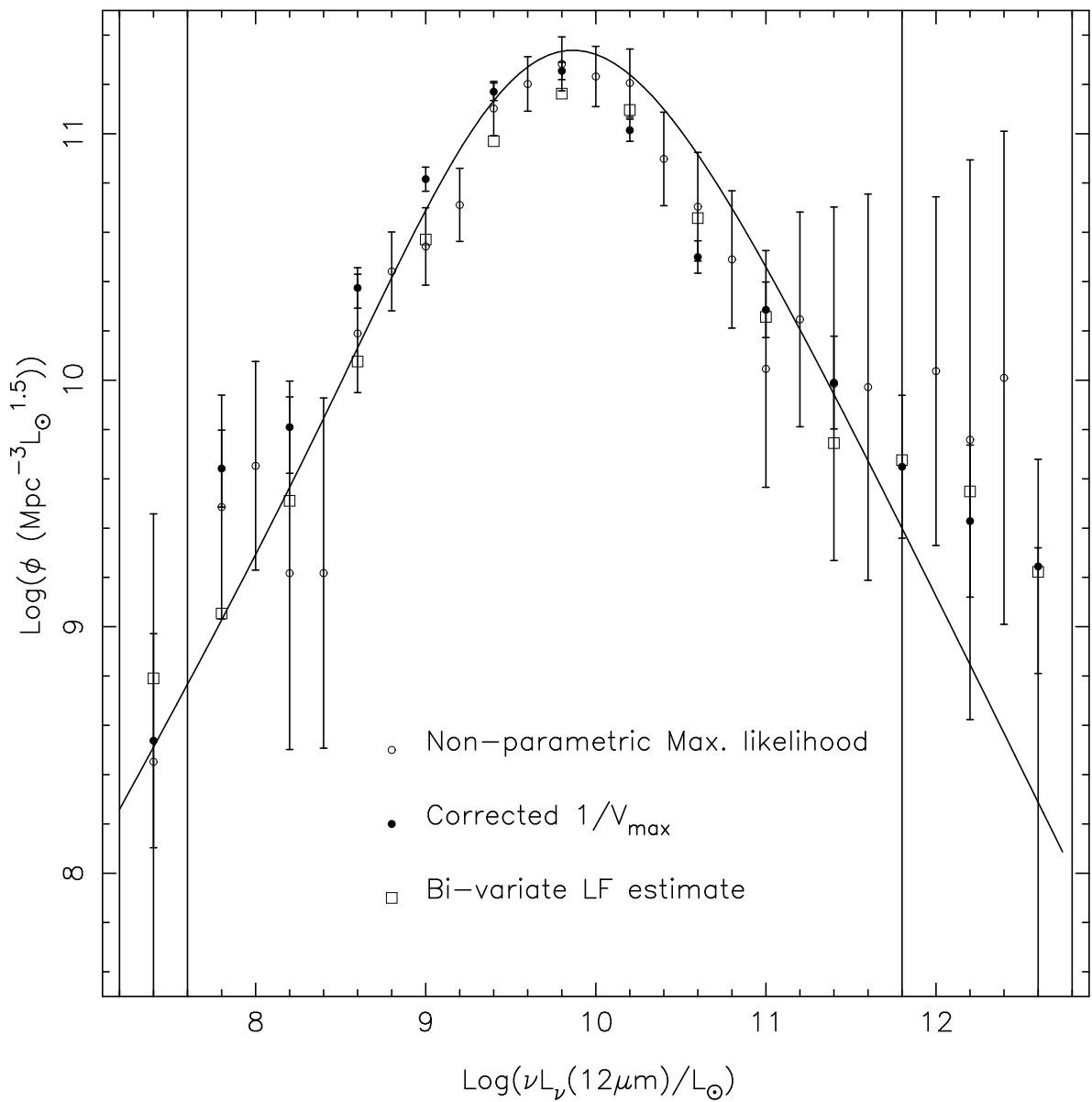}
\caption{
The $1/V_{max}$ 12 $\mu$m visibility function after the density correction
(heavy dots).  It is normalized to the total number of sources in the sample.
Also shown are the maximum-likelihood estimates of Figure \ref{fig:lfrawpah},
and the 1-$\sigma$ error bars.
Squares show the results using a bi-variate function technique which
derives the 12 $\mu$m luminosity function from the 25 $\mu$m one in Paper I.
\label{fig:lfpah}}
\end{figure}

The standard $V/V_{max}$ test (\markcite{schm68}Schmidt 1968)
examines whether the spatial distribution of the sources in a sample
is uniform.  The circles in Figure \ref{fig:vovmax} shows the
result for the sample.  The $1/V_{max}$ estimator is affected by
the variations of the $V/V_{max}$ values shown in the plot.
Following the techniques discussed in Paper I, we have calculated a
radial density distribution for the galaxies in the sample using
the non-parametric maximum-likelihood method.  We have corrected the
$1/V_{max}$ estimator and the $V/V_{max}$ test using the radial
density distribution.  There is marginal improvement on the values of
$V/V_{max}$ (heavy dots) which are now closer to 0.5, shown in
Figure \ref{fig:vovmax}.  The corrected $1/V_{max}$ luminosity function,
normalized to the total number of sources in the sample,
is shown in Figure \ref{fig:lfpah}, where the same maximum-likelihood
results are also plotted.  The visibility function becomes
smoother at low luminosity bins, while the flattening at the high
luminosity bins is somewhat more significant.

We have also derived the 12 $\mu$m luminosity function from the
25 $\mu$m local luminosity function (Paper I) using a bi-variate
luminosity function technique.  Define $\rho_{12}(L_{i})$ and
$\rho_{25}(L_{j})$ as the 12 $\mu$m and 25 $\mu$m luminosity
functions at luminosity bins $L_{i}$ and $L_{j}$ (assuming the same binning
size), respectively,
and $P_{i,j}$ as the conditional
probability of finding galaxies with the luminosity
$L_{i}$ at 12 $\mu$m and the luminosity $L_{j}$ at 25 $\mu$m.  Then
\begin{equation}
\label{eqn:biv}
\rho_{12}(L_{i})=\sum_{j}P_{i,j}\rho_{25}(L_{j}).
\end{equation}
The probability $P_{i,j}$ is calculated empirically (non-parametric)
from the 25 $\mu$m flux-limited sample, taking into account 
also the information content of 12 $\mu$m upper limits using the
Kaplan-Meier estimator (Schmitt 1985; Feigelson and Nelson 1985).
Using the parametric maximum-likelihood
estimate of the 25 $\mu$m luminosity function, which agrees well
with the non-parametric maximum-likelihood and the density-corrected
$1/V_{max}$ results as discussed in Paper I, we obtained the
12 $\mu$m luminosity function, shown by the squares
in Figure \ref{fig:lfpah}.  While the results match the $1/V_{max}$
estimates reasonably well at high luminosity bins, they are
lower at low luminosity bins, agreeing with the maximum-likelihood
estimates.  The greater estimates of the density-corrected $1/V_{max}$
method in these bins are probably due to the left-over effect caused by
structures in the local density distribution, seen in the
density-corrected $V/V_{max}$ test results (heavy dots in Figure
\ref{fig:vovmax}).  Caution should be taken when using the $1/V_{max}$
estimates of the 12 $\mu$m luminosity function at these luminosity bins.
We have obtained the $1/V_{max}$ luminosity function and the $V/V_{max}$
test results calculated for the southern sky subsample (265 galaxies
at $b < -30^\circ$).  The $V/V_{max}$ test shows that 
the subsample is fairly homogeneous, namely the mean $V/V_{max}$ 
are consistent with 0.5 for different flux limits and different $L_{12\mu m}$
bins (see also Paper 1).  Therefore the luminosity function
should not be affected by the local structure significantly.
There is good agreement between the bi-variate luminosity function
and the $1/V_{max}$ luminosity function of the southern subsample,
both are below the density-corrected $1/V_{max}$
luminosity function at low luminosities. 

It should be pointed out that given the small size of the
southern subsample, the luminosity function calculated
from it has relatively large statistical uncertainties.
On the other hand, the bi-variate luminosity function also has larger
error bars due to the additional uncertainties introduced in the
calculation of the probability $P_{i,j}$, in particular
at both ends of the 25 $\mu$m luminosity where the number of
galaxies in each bin is very small.  Potentially, it may also
miss any population which is present in 12 $\mu$m but absent
in 25 $\mu$m, although we do not see any evidence for this.

\begin{figure}
\plotone{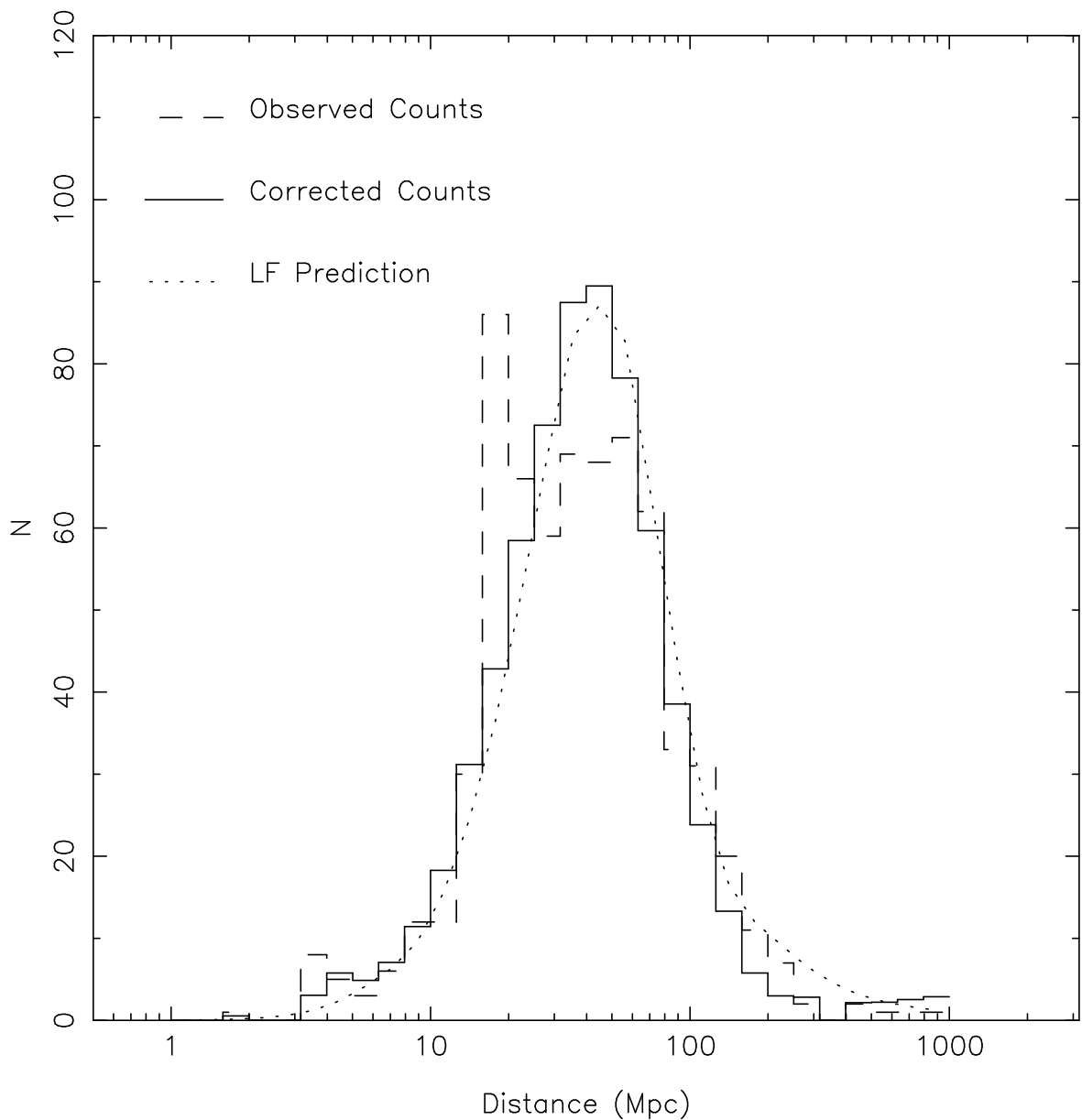}
\caption{
The redshift distribution of the sources in the 12 $\mu$m sample.
The observed distribution (dashed-line histogram) and the distribution
after the radial density correction (solid-line histogram) are shown.
The dotted line is a prediction by the density-corrected $1/V_{max}$
luminosity function in Figure \ref{fig:lfpah}.
\label{fig:dendis}}
\end{figure}

We have compared the redshift distribution expected by the estimated
luminosity function with the one given by the sample.  The observed
redshift distribution can be corrected by the maximum-likelihood
estimate of the radial density distribution to remove the effect of
density inhomogeneities (see Paper I).
In Figure \ref{fig:dendis}, the dashed and solid line histograms show
the distributions before and after the density correction, respectively.
We can see that structures such as Virgo cluster represented by the
peak at $\sim$ 17 Mpc are suppressed and the redshift
distribution becomes smoother after the density correction.
The dotted line is the prediction by the density-corrected $1/V_{max}$
luminosity function, which fits the solid histogram rather well.

The results of the $1/V_{max}$ luminosity functions before and after
the density correction are listed in Table \ref{tab:vmaxlf}.  We
do not include a reliable parametric estimate for the 12 $\mu$m
luminosity function,  since the the flattening at high luminosity bins
is difficult to account for by the current parametric forms.

\subsection{Luminous galaxies at 12 $\mu$m}
\label{sec:nopahlf}

The flattening of the 12 $\mu$m visibility function at high luminosity bins
is unlikely due to the mid-infrared emission features.
We have calculated the visibility function of the 12 $\mu$m continuum from a
subsample drawn from the current sample with the flux density of the
continuum (i.e. the flux density with the mid-infrared emission features
removed) limited at 200 mJy.  The result indicates that the flattening
tail still remains without the emission features.

As discussed in Section \ref{sec:pah}, the strength of the continuum
appears to correlate with that of the nearby emission features.
If they have the same origin, both can be
reduced in regions where radiation intensities are high, in which case
multiple-photon heating of small dust particles
can contribute more significantly to the 12 $\mu$m continuum.
For luminous galaxies with strong radiation,
significant effect on their 12 $\mu$m fluxes from the emission features
is therefore quite unlikely.

The flattening effect is most probably due to the dominance of
a population of Seyfert galaxies and quasars in these bins.
Among the 15 galaxies with $L_{12\mu m} \ge 10^{11}L_{\odot}$, for example,
there are 6 galaxies having the morphology of either a Seyfert 1 or
a Seyfert 2 galaxy, 6 quasars, 1 BL LAC object, and 2 luminous infrared
galaxies.  The Seyfert galaxies and quasars dominate the high luminosity
bins at 12 $\mu$m.  The ultraluminous infrared galaxies are scarce
at these luminosities.  A 12 $\mu$m flux-limited sample selects favorably
the active galaxies and quasars, especially at high luminosities.
It is therefore essential to use longer wavelength bands, such as 25 $\mu$m,
to detect ultraluminous infrared galaxies.
Spinoglio \& Malkan (1989) also showed that the 12 $\mu$m fluxes remain
approximately a constant 25$\%$ of the bolometric flux for all types
of Seyfert galaxies and quasars.  A complete 12 $\mu$m-selected sample
of these galaxies is therefore also complete to some bolometric flux.

\begin{figure}
\plotone{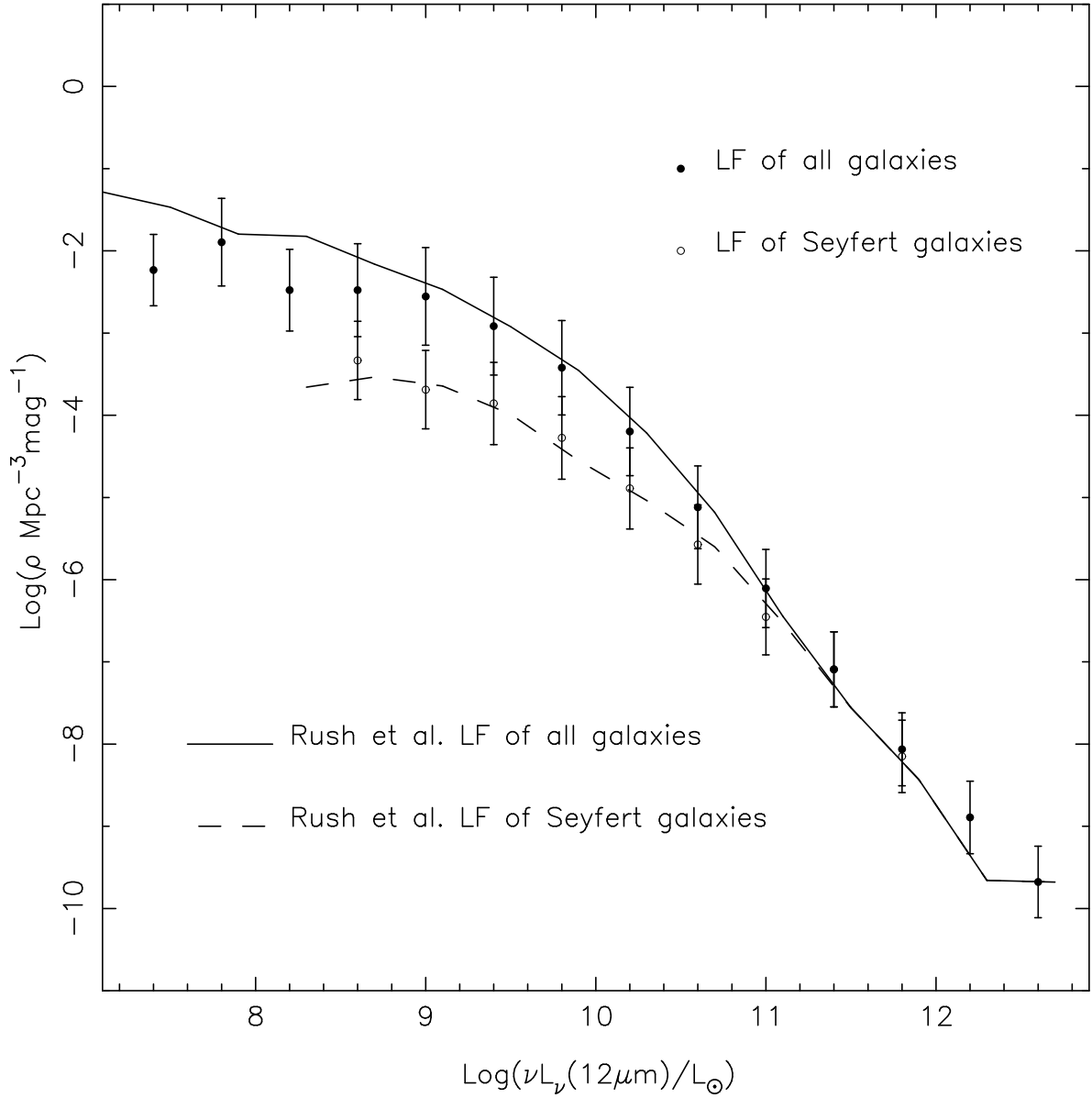}
\caption{
The 12 $\mu$m luminosity functions of all populations of galaxies (heavy dots)
and Seyfert galaxies (circles) of our sample comparing with the ones (lines)
obtained by Rush et al. (1993).  Density variation correction is not performed
for any of the data.  The 1$\sigma$ error bars are shown for our results.
\label{fig:lfcomp}}
\end{figure}

We have identified the active galaxies (including Seyfert galaxies,
BL LACs, and quasars) in our sample using NED.
They form a flux-limited subsample of 109 galaxies.  We obtained
the 12 $\mu$m luminosity function for this subsample, shown by the
circles in Figure \ref{fig:lfcomp}.  Previously, Rush et al. (1993)
obtained the luminosity functions for Seyfert and non-Seyfert galaxies.
Their samples were also 12 $\mu$m-selected but under different criteria.
Figure \ref{fig:lfcomp} shows their 12 $\mu$m luminosity function of
the Seyfert galaxies (dashed line) for comparison.  In the same Figure,
we also compare our $1/V_{max}$ luminosity function (without density
variation correction) for the entire population with
their result.  The Seyfert galaxies in their sample also dominate the
high luminosity bins, shown by the merging of the functions at these bins.
There is good agreement between our results, except at low luminosities for
the samples including all galaxies.  Since the local supercluster can
introduce overestimates of the luminosity function in these luminosity bins,
our density-corrected results should be more reliable, as discussed
in Section \ref{sec:pahlf}.

\section{Summary and conclusions}
\label{sec:conclusion}

Our main results are summarized as follows:

1. Quasars and Seyfert galaxies dominate the high luminosity regime
in a 12 $\mu$m flux-limited sample.  The ultraluminous infrared galaxies
are relatively rare at 12 $\mu$m (contrast to a 25 $\mu$m sample).

2. We have a technique for differentiating between quasars and ultraluminous
infrared galaxies using their 12-25 $\mu$m color (see Figures 5-9).
Qualitatively, quasars are bluer than the luminous and ultraluminous infrared
galaxies at high 25 $\mu$m luminosities.  The ultraluminous infrared galaxies
also have greater far-infrared to blue luminosity ratio on average than those
of the other populations.

3. A highly complete sample flux-density limited at 200 mJy
at 12 $\mu$m selected from the Faint Source Survey catalogs is used
to calculate a local 12 $\mu$m luminosity function, which is then
corrected for density-variation.

We are establishing a library of galaxy SEDs as a function of luminosity
for more accurate $k$-corrections, and will discuss the faint source counts
based on our 12 and 25 $\mu$m luminosity functions in a forthcoming paper
(Xu et al. 1997).

\acknowledgments 

We are grateful to George Helou and Nanyao Lu at IPAC for providing us
with the results of the mid-infrared emission features of 16 galaxies
observed by ISO and the spectra template. 
We thank Carol Lonsdale and Tom Soifer for helpful discussions.
We wish to thank an anonymous referee and our editor, Greg Bothun,
for making helpful comments.
The SIMBAD database is maintained by CDS in Strasbourg, France.  The
NED database is supported at IPAC by NASA.



\clearpage

\clearpage

\begin{deluxetable}{llr}
\tablewidth{0pt}
\tablecaption{Results of the Buckley-James Linear Regression\tablenotemark{1} \label{tab:bj}}
\tablehead{
\colhead{X in $log(F_{25\mu m}/F_{12\mu m})=S\times log(X)+I$}          &   \colhead{slope S}   &  \colhead{intercept I}}
\startdata
$X=\nu L_{25\mu m}$, whole sample        &     0.1556          & -1.1183 \nl
$X=\nu L_{25\mu m}$                      &     0.1543          & -1.0989 \nl
$X=\nu L_{60\mu m}$                      &     0.1587          & -1.2114 \nl
$X=L_{MIR}$                              &     0.1413          & -0.9949 \nl
$X=L_{fir}/L_{b}$                        &     0.2565          & +0.2894 \nl
$X=\nu L_{25\mu m}$, starbursts          &     0.1815          & -1.3560 \nl
$X=\nu L_{25\mu m}$, luminous quasars    &     0.1203          & -1.1595 \nl
$X=L_{fir}/L_{b}$, starbursts            &     0.2657          & +0.2697 \nl
\enddata
\tablenotetext{1}{All relations are calculated using the 25$\mu$m-selected
high-quality sample except for the first one, which uses the entire 25 $\mu$m-selected
sample, and the ones with $L_{fir}/L_{b}$ ratio, for which the blue magnitudes
are obtained for 1390 sources in the entire 25$\mu$m sample.}
\end{deluxetable}

\begin{deluxetable}{ccccc}
\tablewidth{0pt}
\tablecaption{$1/V_{max}$ Luminosity Functions\label{tab:vmaxlf}}
\tablehead{
 & \multicolumn{2}{c}{before density correction} & \multicolumn{2}{c}
{after density correction}  \nl \cline{2-5} \nl
\colhead{log($\nu L_{\nu}$(12$\mu$m)/$L_{\odot}$)}
& \colhead{log($\phi$ Mpc$^{-3}$)}
& \colhead{$1\sigma$ error} & \colhead{log($\phi$ Mpc$^{-3}$)}
& \colhead{$1\sigma$ error}}
\startdata
7.4  & -2.2340    & -2.2340   & -2.4621    & -2.5621  \nl
7.8  & -1.8952    & -2.3251   & -1.9585    & -2.5032  \nl
8.2  & -2.4780    & -2.8358   & -2.3908    & -2.8570  \nl
8.6  & -2.4781    & -3.2257   & -2.4256    & -3.2489  \nl
9.0  & -2.5541    & -3.4929   & -2.5844    & -3.6320  \nl
9.4  & -2.9154    & -3.9942   & -2.8297    & -4.0138  \nl
9.8  & -3.4214    & -4.5240   & -3.3453    & -4.5385  \nl
10.2 & -4.1968    & -5.1940   & -4.1859    & -5.2694  \nl
10.6 & -5.1198    & -5.9324   & -5.3002    & -6.2212  \nl
11.0 & -6.1068    & -6.6860   & -6.1141    & -6.8001  \nl
11.4 & -7.0907    & -7.4510   & -7.0098    & -7.4734  \nl
11.8 & -8.0624    & -8.2356   & -7.9510    & -8.2264  \nl
12.2 & -8.8921    & -9.0408   & -8.7717    & -9.0204  \nl
12.6 & -9.6769    & -9.6769   & -9.5553    & -9.6553  \nl
\enddata
\end{deluxetable}

\end{document}